
\input harvmac
{\catcode`\@=11 \gdef\secn@m.{}}
\Title
{\vbox{\hbox{HUTP--93/A031,}}
       \vbox{\hbox{KCL-TH-93-13}}     }
{\vbox{\centerline{On the Uniqueness of String Theory}}}
\vskip .2in

\centerline{ Nathan Berkovits}
\vskip .2in
\centerline{Math Dept., King's College}
\centerline{Strand, London, WC2R 2LS, United Kingdom}
\vskip .3in
\centerline{ Cumrun Vafa}
\vskip .2in \centerline{Lyman Laboratory of Physics, Harvard University}
\centerline{Cambridge, MA 02138, USA}
\vskip .3in
We show that bosonic strings may be viewed as a particular class of
vacua for $N=1$ superstrings, and $N=1$ superstrings
may be viewed as a particular class of vacua for $N=2$ strings.
Continuing this line of string hierarchies, we are led to search
for a universal string theory which includes all the rest
as a special vacuum selection.

\Date{10/93}

One of the most beautiful aspects of string theory is that it has
almost no adjustable parameters.  The main choice
to be made is the selection of a string vacuum.  The
only non-unique feature seems to be the selection of
which worldsheet symmetries we are gauging.   If we take
the worldsheet gravity theory to be pure gravity (N=0),
we get the
bosonic strings, for $N=1$ supergravity we get
 the fermionic string, and for
$N=2$ supergravity we get the $N=2$ strings.
For closed strings,
we can also choose heterotic combinations $(p,q)$
depending on which symmetries we choose for the left- or right-moving
degrees of freedom (the most well known example being the (0,1) heterotic
string).  In this paper we show that
the choice of which string we consider
 may also be viewed just as another choice of string vacuum.
In particular we will show that {\it any} string vacuum for a $(p,q)$ string
can be viewed as a {\it special choice} of the string vacuum for $(p',q')$
string with $0\leq p\leq p'\leq 2$ and $0\leq q \leq q'\leq 2$.
In other words we have a hierarchy $N=0 \subset N=1 \subset N=2$ and
so in this class, the $(2,2)$ strings is the most general string which
includes all the others as special choices of vacua.

When we speak of a $(p,q)$ string vacuum we
 mean a conformal theory
with $(p,q)$ superconformal symmetry with appropriate central charge\foot{
One can attempt some generalizations of the allowed vacua but we will not
consider them here.  In particular for strings with $c<1$ where the ghost
system is mixed with the matter for  a generic perturbation, we will have
to broaden this strict definition of the vacuum.  We expect that the hierarchy
uncovered in this paper continues to hold even after we broaden the definition
of the vacuum for each string theory.  In this paper we limit
ourselves to the case that all the cohomology comes from the matter
sector with no mixing with the ghost.}:
$(N=0\rightarrow c=26; N=1\rightarrow {\hat c}=10;
N=2\rightarrow {\hat c}=2)$, where
the last $\hat c=2$ corresponds to
counting complex dimensions.  What we shall say below for the left-movers
can also be said for the right-movers, so in the following we
will limit our discussion to the left-movers.

\subsec{$N=0 \subset N=1$}
Consider a bosonic string vacuum, i.e. a conformal theory
with $c=26$.  It is known \ref\super{B. Gato-Rivera
and A.M. Semikhatov, Phys. Lett. B293 (1992) 72-80\semi
M. Bershadsky, W. Lerche, D. Nemeschansky, N.P. Warner,
{\it Extended N=2 Superconformal Structure of Gravity and W Gravity
Coupled to Matter} CERN-TH-6694-92, hep-th/9211040 .}\
that if we use a $U(1)$ current of the matter system, there
is a hidden $N=2$ supersymmetry in the combined
matter-ghost system.  However we have to assume the existence
of a current in the matter system. In particular, choosing such
a current in the case of strings on $R^{26}$ would break
Lorentz invariance.  However it turns out that if we slightly
change the spin content of the ghost system (roughly speaking
`untwist it')
without choosing any $U(1)$ currents, there is an
$N=1$ supersymmetry for the combined matter and shifted spin
$(b,c)$ ghost system.  Let us call the shifted spin ghost
system by $(b_1,c_1)$ which has spin (3/2,-1/2).  Then denoting
the energy momentum tensor for the matter system as $T_m$ (with
central charge 26), we can write the generator of the $N=1$
superconformal alegebra as
$$G=b_1+c_1(T_m+\partial c_1 b_1)+{5\over 2}\partial^2 c_1$$
\eqn\bos{T=T_m-{3\over 2}b_1\partial c_1-{1\over 2}\partial b_1 c_1+{1\over 2}
\partial^2(c_1\partial c_1) .}
We have the following OPE:
$$G(z)G(0)\sim {10\over z^3}+{2T\over z}$$
\eqn\bosi{T(z)T(0)\sim {15\over 2z^4 }+{2T\over z^2}+{\partial T\over z}}
which is an $N=1$ superconformal algebra with $\hat c=10$
($c=15$).  Note that the energy-momentum tensor is the
sum of the energy momentum tensor of the matter system and
the spin (3/2,-1/2)-system  $(b_1,c_1)$, with an extra `improvement'
term which can be written as $\partial J/2 $ where
$J=\partial (c_1\partial c_1)$.  Note that since $J J\sim 0$,
it does not affect the central charge of the system.
In the path-integral
language, this is the same as adding a term proportional
to $\int R (c_1 \partial c_1)$ to the action. Since
this term violates $c_1b_1$ number by 2 units, it will not
contribute to correlations that we shall consider below
due to $c_1b_1$ ghost number conservation.

The existence of an $N=1 $ superconformal
algebra in the bosonic string after shifting the spins of the
ghosts is non-trivial and is not the consequence of the $N=2$
superconformal symmetry considered in \super \foot{
In particular the map considered there takes the critical bosonic
string on $R^{26}$ to the N=2 conformal
theory with $\hat c=3$ ($c=9$).}.  However the form of
$G$ is reminiscent of $G^-+G^+$ from that reference where
$G^-=b$ and $G^+=j_{BRST}$ (up to a total derivative term).
Note also that for $G$ to have spin $3/2$ we needed to shift
the spins of $(b_1,c_1)$ to be (3/2,-1/2).

Thus we see that the combined system $(T_m, (b_1,c_1))$
has $N=1$ superconformal symmetry with ${\hat c}=10$ which
thus makes it a viable candidate for the {\it matter system} of an
$N=1$ fermionic string.  We will now show the equivalence
$$(T_m,(b_1,c_1);(b,c),(\beta ,\gamma ))\rightarrow
(T_m;(b,c))$$
where the lefthand side refers to an $N=1$ vacuum
with $((b,c),(\beta,\gamma))$ denoting the superdiffeomorphism
ghosts.  Formally it seems plausible that this correspondence
might work since if
we ignore the zero modes
of the fields, on the lefthand side the
fermionic $(b_1,c_1)$ system
cancels the bosonic ghost $(\beta ,\gamma)$
because they have the same spin but opposite statistics.
We are thus left with $T_m$ and the $(b,c)$ ghosts,
which is all that we have on the righthand side.
Before going on to show how the zero modes work,
let us discuss the physical fields of the resulting string theory
and the computation of amplitudes at the tree level.

Bosonic string states are described by dimension 1
vertex operators, $V$, which will be assumed to be constructed entirely
out of the matter fields in $T_m$. Although $c_1 V$ commutes with the
BRST charge of the bosonic string, it is a dimension $1/2$ operator after
shifting the spin of the $c_1$ field. The corresponding BRST-invariant
operator for the N=1 string is therefore $c e^{- \phi} c_1 V$
(this is $N=1$ BRST invariant since
$c_1V$ is a primary of the $N=1$ superconformal matter system
with the appropriate dimension).  In N=1
language, this operator is in the ghost-number 0 picture and to get other
pictures, one attaches the picture-changing operator
\eqn\pic{Z=\{ Q_{N=1}, \xi \}= e^{\phi}G + 2\partial\eta e^{2\phi} b +
\eta \partial(e^{2\phi} b) +c\partial\xi }
where $G$ is defined in equation (1)
and the ($\xi,\eta,\phi$) fields come from
bosonizing the $(\beta,\gamma)$ ghosts in the usual way. For example,
the BRST-invariant
operator in the ghost-number 1 picture is $c V + \gamma c_1 V$.
The integrated form of these operators is
obtained in the usual way by commuting with $\int b$ which leads
to $\int V$.

For n-point amplitudes on a sphere, the relevant correlation function is
therefore:

\eqn\bost{\langle (c\ e^{-\phi}c_1V_1)  (c\ e^{-\phi}c_1V_2) (c\ V_3 +\gamma
c_1 V_3)
 \int V_4 ...\int V_n
\rangle }
Note that the second part of the vertex operator for $V_3$
does not contribute due to $b_1 c_1$ conservation (as mentioned earlier,
this conservation also allows us to ignore the `improvement' term
$\partial(c_1\partial c_1)$ that appeared in $T$ of equation
\bos ). After taking
into account the extra $c_1 e^{-\phi}$
that appear in the first two vertex
operators, it is
easily checked that the path integral over the $(b_1,c_1)$ fields
precisely cancels the path integral over the $(\beta,\gamma)$,
including the zero modes, leaving us with
the usual bosonic string correlation function for the tree amplitude.

For surfaces of genus g, the relevant N=1 correlation function is
$$\langle
  \prod_{j=1}^{3g-3}\int dy_j \mu_j(y_j) b(y_j)
\quad \prod_{k=1}^{2g-2} Z(u_k)
\quad \int V_1 ...\int V_n
\rangle$$
where $Z$ is the picture-changing operator
defined in equation \pic\ and $\mu_j$ comes from the
usual $3g-3$ Beltrami differentials. By $b_1 c_1$ conservation, only the
$e^{\phi} b_1$ term from $Z$ contributes to the correlation function (the
other terms in $Z$ all have non-positive $b_1 c_1$ number and the total
$b_1 c_1$ number must equal $2g-2$).
Since $e^{\phi} b_1=\delta(\beta )b_1$
\ref\ver{ E.Verlinde and H.Verlinde, Phys. Lett. B192 (1987) 95.},
the path integral
over the $(b_1,c_1)$ fields again precisely cancels the path integral
over the $(\beta,\gamma)$ fields, including the zero
modes,
and we are left with the usual bosonic
string correlation function on a genus $g$ surface. Note that this
correlation function is independent both of the position of the $Z$'s
and of the spin structure chosen for the fields with half-integer spin.

However there is a point that we need to be more careful about.
In the superstring theory, we are instructed to sum
over all spin structures and divide the resulting amplitude
by $2^g$.  We have just argued that for each spin structure,
the resulting amplitude is the same as that for the bosonic
string.  Therefore summing over all spin structures
would give us a factor
$$\big[ {1\over 2^g} \sum_{{\rm spin\ structures}}1\big] Z_{bosonic}
={2^{2g}\over 2^g}Z_{bosonic}=2^g Z_{bosonic}$$
But this is in contradiction with the unitarity of the fermionic string.
We know that the bosonic string is unitary, and moreover, the only
way to modify the genus $g$ partition function consistent
with unitarity is to let $Z_g\rightarrow \lambda^{g-1} Z_g$.
But the above factor is $2^g$ and not $2^{g-1}$, which means that
we cannot absorb it into a redefinition of the coupling constant.
The resolution of this puzzle follows from the fact that
we know only that the norm of the contribution from
each spin structure is the same as that of the bosonic string.
We cannot argue about its phase.  However invariance under modular
transformations (which mix even and odd spin-structures seperately)
implies
that we have essentially only two choices.  Let us assume
that all the even spin structures come with a factor of (+1)
and all the odd ones with a factor of (-1).  We thus see that
the computation of the $N=1$ side leads to
$${1\over 2^g}\big[ \sum_{\rm spin\ struct.}(-1)^\sigma \big]Z_{bosonic}
={1\over 2^g}[2^{g-1}(2^g+1)-2^{g-1}(2^g-1)]Z_{bosonic}=Z_{bosonic}$$
This is exactly what we wished to have with no adjustments
of coupling constants!  So it is crucial for the consistency of our
equivalence that the difference between the number of even
and odd spin structure in genus $g$ be $2^g$, which magically enough
is the case!

If we apply this computation to the genus 1 case, what we are saying
is that instead of getting ${1\over 2}(1+1+1+1)=2$ times the bosonic
string amplitude, we get the three even spin structures contributing
a $+1$ and the odd one contributing a (-1): ${1\over 2} (1+1+1-1)=1$.
Translating this to the language of the Hilbert space, we see that
the Ramond sector does not contribute any physical states
for this vacuum because all the states
in it are GSO odd!  So all the physical states come from the
NS sector and they are in one to one correspondence with the
bosonic string states.

This concludes
showing that we may
imbed the $N=0$ string in the $N=1$ string, albeit as a very
particular subclass of $N=1$ vacua.

\subsec{$N=1\subset N=2$}
We can now proceed
by analogy to imbed the $N=1$ string in the $N=2$ string.
The idea is to start with an $N=1$ SCFT with $\hat c=10$, add the
usual $N=1$ ghosts, then shift the spins of the ghosts down by $1/2$ and
try to construct out of this total system an $N=2$ SCFT with $\hat c=2$
(where $\hat c =2 $ refers to the $N=2$ central charge, i.e., it corresponds
to $c=6$).  Then we should take this system as the matter system for
the $N=2$ strings, add the $N=2$ ghost system, and show that the
computations
of amplitudes in the $N=2$ formalism with this matter system is the
same as in the original $N=1$ SCFT coupled to $N=1$ supergravity.

Let us represent the $N=1$ matter system with $\hat c=10$
by $(T_m,G_m)$,
i.e. by its energy momentum tensor and by its supercurrent.
Let us denote the spin-shifted ghost system by
fermionic $(b_1,c_1)$ of spin (3/2,-1/2) and bosonic $(\beta_1,\gamma_1)$
of spin (1,0).  Then the claim is that this system magically has
an $N=2$ superconformal symmetry with $\hat c=2$.
This same $N=2$ algebra was constructed in \ref\ber{N. Berkovits,
{\it The Ten Dimensional Green-Schwarz Superstring is a Twisted
Neveu-Schwarz-Ramond String},hep-th/9308129.}\
in showing that $N=1$ NSR strings can be mapped to GS strings
in the $N=2$ formulation.
In the following we will sometimes find it convenient to use the bosonized
version of the $(\beta_1,\gamma_1)$ system, in which case we write
$$\gamma_1=\eta_1 {\rm exp}(\phi)\qquad \beta_1=\partial \xi_1{\rm
exp}(-\phi )$$
where $\phi$ is the usual bosonized ghost field of the fermionic
string \ref\fms{D. Friedan, E. Martinec and S. Shenker,
Nucl. Phys. B271 (1986) 93.}\ with appropriate background charge and
the $(\eta_1 ,\xi_1)$ are spin-shifted to (3/2,-1/2) in order
to reproduce correctly the spin-shifted $(\beta_1 ,\gamma_1)$ system
(or more appropriately, we could have started with the bosonized version
of this system as the fields to be added
to the matter system). Now we are ready to write the $N=2$ generators
of the combined system $((T_m,G_m),(b_1,c_1),(\beta_1 ,\gamma_1))$:
$$G^-=b_1$$
$$ G^+=\gamma_1G_m+c_1(T_m-{3\over 2}\beta_1
\partial \gamma_1 -{1\over 2}\partial \beta_1 \gamma_1)-
\gamma_1^2b_1 +\partial (c_1\xi_1 \eta_1)+\partial^2 c_1+b_1c_1
\partial c_1$$
$$T=T_m-{3\over 2}\beta_1 \partial \gamma_1 -
{1\over 2}\partial \beta_1 \gamma_1 -b_1\partial c_1 -{1\over 2}
\partial(b_1c_1 -\xi_1\eta_1)$$
\eqn\ferm{J=b_1c_1+\xi_1\eta_1}
It is tedious but a straightforward exercise to show that these
form an $N=2$ superconfomal system with $\hat c=2$, which is the critical
value needed for the N=2 string.
Note in particular that we could not write the $N=2$ algebra
in a simple form just in terms of the
$\beta_1 ,\gamma_1 $ system because of the
appearance of $\xi_1\eta_1$ in the currents. Furthermore, it does not
seem possible to enlarge this algebra to the N=3 algebra found in
reference [1] since $\beta_1$ no longer has the right OPE with $J$
to be an $SO(3)$ current.

We now take this matter system and couple it to $N=2$ supergravity.
So the total system including the ghosts is
$$((T_m,G_m),(b_1,c_1),(\eta_1,\xi_1),\phi;(b,c),(\beta^{\pm} ,\gamma^{\pm}),
(\eta ,\xi ))$$
where $(b,c)$ are the fermionic diffeomorphism
 ghosts of spin (2,-1), $(\beta^{\pm}
\gamma^{\pm})$ are the bosonic superdiffeomorphism ghosts of spin (3/2,-1/2),
and $(\eta ,\xi )$ are the fermionic $U(1)$ diffeomorphism ghosts of spin
(1,0).
We will now show that scattering amplitudes calculated using this
N=2 string theory agree
with amplitudes calculated using
the original $(T_m,G_m)$ system coupled to N=1 supergravity.

First let us argue how the non-zero modes work as before:  Consider
an arbitrary $N=2$ moduli, which consists of a worldsheet moduli
and the choice of a flat $U(1)$ connection.  Note that the
only $U(1)$ charged fields are
$((b_1,c_1),(\eta_1,\xi_1);(\beta^{\pm },\gamma^{\pm}))$
which all have spin (3/2,-1/2),
and the first two are fermionic while the last two are bosonic.
Thus the non-zero modes
of all of these fields cancel out and we are left with
$$((T_m,G_m),\phi;(b,c),(\eta ,\xi))$$
which is precisely the matter and ghost content of the $N=1$ string
(when we think of the leftover $(\phi ,\eta ,\xi)$ as the bosonized
$(\beta ,\gamma)$ ghost).  Before describing how the zero modes
work, let us construct the physical vertex operators of the resulting
$N=2$ string theory and compute the tree level amplitudes.

In N=1 language, physical states are described by vertex operators in
different pictures. For example, Neveu-Schwarz states can be described by the
operator $c_1 e^{-\phi} V$ in the ghost-number 0 picture or by the
operator $c_1 [G_m,V] + \gamma_1 V$ in the ghost-number 1 picture.
In the above expressions,
$V$ is assumed to be a dimension 1/2 operator constructed
entirely out of N=1 matter fields and $[G_m,V]$ means the $z^{-1}$
piece from the OPE of $G_m$ with $V$.
Similarly, Ramond states can be described by the operator
$c_1 e^{-\phi /2} W$ in the ghost-number 1/2 picture or by
$c_1 e^{+\phi /2} [G_m ,W] +(b_1 c_1 \eta_1+2\partial\phi \eta_1
+2\partial \eta_1) e^{3\phi /2} W$ in the ghost-number 3/2 picture,
where
$W$ is a dimension 5/8 operator constructed
entirely out of N=1 matter fields and $[G_m ,W]$ means the $z^{-1/2}$
piece from the OPE of $G_m$ with $W$.

In N=2 language, there are not only two picture-changing operators
defined by
\eqn\pict{Z^\pm= [Q_{N=2},\xi^\pm]=e^{\phi^\mp}[G^\pm +(b\mp 1/2
\partial\eta)\gamma^\pm
-\eta\partial\gamma^\pm ]+c\partial\xi^\pm ,}
but also instanton-number-changing operators
\foot{The instanton charge
is defined as $N_I=\int (\partial\phi -\eta_1\xi_1)$ which is N=2 BRST
invariant and
satisfies $[N_I,I^\pm ]=
\pm I^\pm$. The instanton number, $n_I$, of an operator $V$ is defined
by $[N_I,V]=n_I V$. If $V$ is U(1)-invariant and is constructed entirely
out of N=2 matter fields, $n_I$ is equal to the N=1 ghost number since
$N_I=\int (c_1 b_1 +\partial\phi +J)$.}
defined by
\eqn\ins{I=e^{\int J +\phi^+ -\phi^-}
=b_1 \xi_1 e^{\phi^+ -\phi^-} \quad {and} \quad
I^{-1}=e^{-(\int J +\phi^+ -\phi^-)}=c_1 \eta_1 e^{\phi^- -\phi^+}}
where $G^{\pm}$ and $J$ are defined in equation
\ferm\ .
It is easy to verify that
these operators are in the N=2 BRST cohomology but their derivatives are
BRST trivial, so they can be used to express the
same physical state in terms of different N=2 vertex
operators. Note that unlike in the N=1 formalism, these
different vertex operators may
contain the same N=2 ghost number. \ref\berk{N. Berkovits, Nucl. Phys.
B395 (1993) 77, hep-th/9208035;
Phys. Lett. B300 (1993) 53  and {\it
Finiteness and Unitarity of Lorentz-Covariant
Green-Schwarz Superstring Amplitudes}, KCL-TH-93-6, hep-th/9303122
(to appear in Nucl.Phys.B).}

For example, a physical Neveu-Schwarz state with N=2 ghost number
$-1$ can be described
by the BRST-invariant vertex operator $c e^{-\phi^+ -\phi^-}\xi_1
c_1 e^{-\phi} V$. A BRST-invariant
ghost-number +1 vertex operator can be constructed
by attaching $Z^- Z^- I^{-1}$, which results in the operator
$c e^{-\phi} V + \gamma^- c_1 e^{-\phi} V$.
Alternatively, one could attach $Z^+ Z^-$,
which results in the BRST-invariant ghost-number +1 vertex operator
$c [G,V] + c\partial(c_1 \xi_1 e^{-\phi} V)+\gamma^-\xi_1 e^{-\phi}V$.

Similarly, a physical Ramond state with N=2 ghost-number 0
is described by the vertex operator
$c e^{-\phi^+ } c_1 e^{-\phi /2} W$.
Ghost-number +1 operators can be obtained by attaching $Z^-$
to get $c e^{-\phi /2} W+\gamma^- c_1 e^{-\phi /2} W$, or by
attaching $Z^+ I$ to get
$c (e^{\phi/2}[G,W]+ \eta_1 b_1 e^{3\phi/2} W +\partial(c_1 \xi_1
e^{-\phi/2} W)) +\gamma^- \xi_1 e^{-\phi/2} W$.

These vertex operators can be expressed in integrated form by
commuting with $\int b$, and it is easy to show that in this form,
they precisely coincide with the integrated form of the N=1 expressions.
For example, the  N=2 Neveu-Schwarz vertex operators take the form
$\int e^{-\phi}V$ and $\int [G,V]$, while the N=2 Ramond vertex operators
take the form $\int e^{-\phi/2}W$ and $\int (e^{\phi/2}[G,W]
+e^{3\phi/2} \eta_1 b_1 W)$. The equivalence of N=1 and
N=2 vertex operators in integrated form is not so surprising since
$[Q_{N=1},V]=0$ and $[b_1,V]=0$ is almost enough
to imply that $[Q_{N=2},V]=0$. The only obstacle may come from singularities
which are total
derivatives in the OPE of $G^+$ with $V$ (these singularities would spoil
the N=2 BRST invariance of $V$, but not the N=1 BRST invariance).
Singularities of this type do not occur in the pictures
considered above, however they probably do occur for vertex
operators in other pictures.

To calculate tree amplitudes with $m$ bosons and
$2n$ fermions in the N=2 formalism, one needs to evaluate
the correlation function
$$\langle  \xi(z_0) (c e^{-\phi^+ -\phi^-}\xi_1 c_1 e^{-\phi} V_1)
(c e^{-\phi^+ -\phi^-}\xi_1
 c_1 e^{-\phi} V_2)
\quad \int [G,V_3] ...\int [G,V_m]$$
$$ (c e^{-\phi/2}W_1+\gamma^- c_1 e^{-\phi/2}W_1)
\int e^{-\phi/2}W_2 ...\int e^{-\phi/2}W_n]$$
$$ \int (e^{\phi/2}[G,W_{n+1}]+e^{3\phi/2}\eta_1 b_1 W_{n+1})
... \int (e^{\phi/2}[G,W_{2n}]+e^{3\phi/2}\eta_1 b_1 W_{2n})
\rangle$$
where the zero mode of the $\xi$ ghost needs to be inserted in order
to get a non-zero amplitude (the correlation function is independent
of the location of the insertion). Because of $e^{\phi}$ conservation,
the second term in the vertex operators $W_{n+1}$ to $W_{2n}$
never contributes to the correlation function. Similarly,
the second term in the vertex operator for $W_1$ does not
contribute by $bc$ conservation. For these reasons, the
$(b_1, c_1)$ and $(\xi_1,\eta_1)$ path integrals are easily shown
to cancel the $(\beta^\pm,\gamma^\mp)$ path integrals. The resulting
correlation function is precisely the relevant one for the N=1
calculation of the scattering amplitude (the second term in
$W_{n+1}$ to $W_{2n}$ can also be ignored in the N=1 calculation because
of $e^{\phi}$ conservation).
Note that the
ability to ignore the second term in $W_{n+1}$ to $W_{2n}$
depended crucially on
the pictures chosen for the other vertex operators, and for a general
choice of pictures, it is not obvious how to prove directly
that the N=1 and N=2 correlation
functions coincide. Of course, they can be proven indirectly
to coincide by using the fact that all pictures are related to each
other by BRST-trivial operations, which only changes the integrand
of the scattering amplitude by a total derivative.

To prove the equivalence for multiloop amplitudes, it is also useful
to choose special pictures for the external states. For a genus $g$
surface, the simplest amplitude to compare contains $m$ Neveu-Schwarz
states, all in the ghost-number 0 picture, and $4g-4+2n$ Ramond states,
$n$ in the ghost-number $-1/2$ picture, and $4g-4+n$ in the ghost-number
$+1/2$ picture. With this choice of picture, there is no need in the
N=1 formalism to insert picture-changing operators at additional points.

In the N=2 formalism, the relevant measure on the moduli space is:
$$\prod_{i=1}^g \int dm_i \langle  \xi(z_0)
\int dw_i\eta(w_i)\quad
  \prod_{j=1}^{3g-3}\int dy_j \mu_j(y_j) b(y_j)\quad
 \prod_{k=1}^{4g-4} Z^-(u_k)$$
$$ \quad \int [G,V_1] ...\int [G,V_m]
\int e^{-\phi/2}W_1 ...\int e^{-\phi/2}W_n$$
$$\int e^{\phi/2}[G,W_{n+1}]+e^{3\phi/2}\eta_1 b_1 W_{n+1}
...\int e^{\phi/2}[G,W_{4g-4+2n}]+e^{3\phi/2}\eta_1 b_1 W_{4g-4+2n}
\rangle$$
where $m_i$ are the complex U(1) moduli whose integration region is
$C^g/(Z^g+\tau Z^g)$,
$\int dw_i\eta(w_i)$ are the corresponding $U(1)$ ghost
insertions, and the $4g-4$ $Z^-$'s
come from integrating over the supermoduli. Note that because the combined
instanton number of the vertex operators is $2g-2$ (it is equal to the N=1
ghost number of the vertex operators), the surface itself must carry
instanton number $2-2g$ to get a non-zero amplitude. On such a surface,
the conformal weight of all fields is shifted
(if one thinks of identifying the spin connection with the
$U(1)$ connection) by their
U(1) charge (i.e. the screening charge for $[\int b_1 c_1,\int \eta_1 \xi_1,
\phi^+,\phi^-]$ is shifted from $[2,2,2,2]$ to $[4,0,4,0]$), and there
are $4g-4$ super-moduli for one of the two right-moving
gravitinos and none for the other
one.

As in the tree amplitude, the second term in the vertex operators for
$W_{n+1}$ to $W_{4g-4+2n}$ can be ignored due to $e^{\phi}$ conservation,
implying by $b_1 c_1$ conservation that only the $e^{\phi^+} b_1$ part of
$Z^-$ contributes to the correlation function (because of the shift in
screening charge, $4g-4$ $b_1$'s are needed for a non-zero result).
Using these facts, it is again easy to show that the $(b_1,c_1)$
and $(\eta_1,\xi_1)$ path integrals precisely cancel the
$(\beta^\mp,\gamma^\pm)$ path integrals
including the zero modes.
Since all other fields contribute identically in
the two formalisms, the integrands of the scattering amplitudes calculated
using the N=1 and N=2 theory coincide\foot{It is important to
study precisely how the left- and right-movers are put together
(in particular the chiral one we are discussing here is
naturally identified with the $Z_\infty$ string of
\ref\ov{H. Ooguri and C. Vafa,
Nucl. Phys. B367 (1991)83 ; B361 (1991) 469;
Mod. Phys. Lett. A5 (1990) 1389.}).  A related issue is that the
integration over the $U(1)$ Jacobian
leaves us with a factor of ${\rm det}\tau_2$ which should be
canceled by an inverse factor from the $\beta_1,\gamma_1 $
part of the matter system as would be dictated by modular invariance.
    Another related issue is the fact
that $\phi$ is a negative-energy boson which may introduce
unphysical poles, as is familiar from the $N=1$ superstrings.
Similar issues arise and have been dealt with in the context
of the N=2 formulation of the Green-Schwarz superstring \berk .}.
Note that because the N=2 matter system has not yet been GSO
projected, one still needs to sum over spin structures in the N=2
matter fields in order to recover the complete N=1 scattering amplitude.

If a different picture is chosen or if
less than $4g-4$ Ramond states are present, it is still expected by unitarity
arguments
that the scattering amplitudes coincide, however it is probably not true
that the integrands of the amplitudes coincide (they only need to agree up
to total derivatives in the moduli). This is not
surprising since although the integrands are required by unitarity to coincide
in light-cone gauge, gauge transformations away from light-cone gauge
may shift the integrands by total derivatives. Since there is no simple
relation between an N=1 gauge transformation and an N=2 gauge transformation
($G$ of N=1 is related in a complicated way to $G^\pm$ of N=2), there
is no reason why the shift of the N=1 integrand and the shift of the N=2
integrand should be equivalent.

\subsec{Conclusions}
We have seen that strings are more unique than had been suspected.
The most general string studied so far, which includes all
the other ones as special cases is the $N=2$ string. It
is rather amusing that the more conventional choice for
the vacuum of the $N=2$ string was also discovered
to have magical properties and in particular be related to
self-dual geometries in four dimensions \ov , and
was conjectured to be the master theory for all
integrable models (see \ref\mar{ N. Marcus, {\it A Tour Through
N=2 Strings}, hep-th/9211059.}\ for a review of the
literature on $N=2$ strings).

However it is natural to ask if the story stops here. In particular,
it has been recently found that the $N=2$ string has a hidden
new $N=4$ superconformal symmetry \ref\rocg{
A. Giveon and M. Rocek, {\it On the BRST Operator Structure
of the N=2 String}, hep-th/9302049
\semi
J. Gomis and H. Suzuki, Phys. Lett. B278 (1992) 266.}\
which suggests that this chain of construction
can be continued at least one more step to the $N=4$ strings
which magically enough has critical dimension zero. Since
the $N=4$ strings have not been studied in depth, this remains conjectural
but we find it aesthetically compelling for the chain
of hierarchy to continue to the $N=4$ case.

There may be other directions in which one can generalize these
constructions:  We have seen that
 there is a hierarchy of strings, depending on which features
of moduli space of Riemann surfaces we decide to focus on.  The more
`general' string theory is the one which allows the partition function to
depend non-trivially on more extra data available. For example,
we have seen that special
vacua of $N=1$ strings that do not depend on spin structure
reduce to the $N=0$ strings. In this sense it is
easy to understand why $N=0$ strings
should be viewed as a special case of $N=1$ strings and not the other
way around. Similarly, special $N=2$ string vacua which
are insensitive to the $U(1)$ moduli of $N=2$ reduce
to the $N=1$ string.  Therefore we are naturally led to ask
what other structures can we put on a Riemann surface?  One
such direction could be $W_N$ strings
(see \ref\pow{C.N. Pope,
{\it W-Strings 93}, hep-th/9309125
\semi P. West,{\it
A Review of W Strings}, hep-th/9309095, Goteborg-ITP-93-40.}\
for a review of recent results), or following
the notion that the most structured
string is the most general one,
we could go on to $W_\infty$ strings
(or some supersymmetric generalization thereof).   It is quite
amusing in this connection that in \ov\ it was
noted that the `most symmetrical' choice for a target space of $N=2$ strings
possesses $W_{\infty}$ symmetry in the target space which
by worldsheet--target duality was conjectured to transmute
to the worldsheet.
It would be interesting to pursue such directions more seriously
and see if we are close to identifying the most general string theory.

At any rate, the search for {\it the} universal string theory (UST),
the one which includes all the others
by special choices of vacua, is now on!
  One would naturally expect that string symmetries
are most manifest in such a universal theory.
The world around us has decided to choose a highly
asymmetrical vacuum of this very symmetrical string,
maybe in the form of the heterotic string.  But,
as is the case with spontaneously broken symmetries,
we expect that for deepest insights into a theory
 we have to understand the most symmetrical
formulation of it and not concentrate on the `asymmetrical'
string vacuum!  In this sense, more or less realistic string
vacua such as the heterotic string play the role
of a `symmetry broken phase' of the UST.

We would like to thank the hospitality of Rutgers University
where this project was initiated. We would also like to thank
Michael Bershadsky, Mike Douglas,
Mike Freeman, Martin Rocek, Nathan Seiberg,
Warren Siegel, and Peter West for useful
discussions.  N.B. would also
like to thank Harvard University for its hospitality, and the SERC
for financial support.
The research of C.V. is supported in part by a Packard
fellowship and NSF grants PHY-87-14654 and PHY-89-57162.
\listrefs

\end